\documentclass[a4paper,UKenglish, 11pt]{article}

\usepackage{amstext,amsmath,amsthm,amssymb,amscd}
\usepackage{xspace}
\usepackage{bm}
\usepackage{paralist}
\usepackage{authblk}
\usepackage{fullpage}
\title{Preprocessing under uncertainty}

\author{Stefan Fafianie}
\author{Stefan Kratsch}
\author{Vuong Anh Quyen}
\affil{University of Bonn, Germany, \{fafianie,kratsch,vuong\}@cs.uni-bonn.de}

\theoremstyle{definition}
\newtheorem{theorem}{Theorem}
\newtheorem{definition}{Definition}
\newtheorem{lemma}{Lemma}

\newtheorem{observation}{Observation}

\newcommand{\Oh}{\mathcal{O}}
\newcommand{\Pee}{\mathcal{P}}
\newcommand{\N}{\mathbb{N}}
\newcommand{\M}{\mathcal{M}}
\newcommand{\I}{\mathcal{I}}
\newcommand{\R}{\mathbb{R}}
\newcommand{\MSF}{\textsc{MSF}}
\newcommand{\MST}{\textsc{MST}}

\DeclareMathOperator{\tor}{torso}
\newcommand{\problem}[1]{\lowercase{\textsc{#1}}}
\newcommand{\mst}{\problem{Minimum Spanning Tree}\xspace}
\newcommand{\mwmb}{\problem{Minimum Weight Matroid Basis}\xspace}
\newcommand{\mcbm}{\problem{Maximum Cardinality Bipartite Matching}\xspace}

\newcommand{\NP}{\ensuremath{\mathsf{NP}}\xspace}
\renewcommand{\P}{\ensuremath{\mathsf{P}}\xspace}

\usepackage{comment}
\specialcomment{shortproof}{\begin{proof}}{\end{proof}}
\specialcomment{longproof}{\begin{proof}}{\end{proof}}

\begin{document}

\maketitle

\begin{abstract}
In this work we study preprocessing for tractable problems when part of the input is unknown or uncertain. This comes up naturally if, e.g., the load of some machines or the congestion of some roads is not known far enough in advance, or if we have to regularly solve a problem over instances that are largely similar, e.g., daily airport scheduling with few charter flights. Unlike robust optimization, which also studies settings like this, our goal lies not in computing solutions that are (approximately) good for every instantiation. Rather, we seek to preprocess the known parts of the input, to speed up finding an optimal solution once the missing data is known.

We present efficient algorithms that given an instance with partially uncertain input generate an instance of size polynomial in the amount of uncertain data that is equivalent for every instantiation of the unknown part. Concretely, we obtain such algorithms for \mst, \mwmb, and \mcbm, where respectively the weight of edges, weight of elements, and the availability of vertices is unknown for part of the input. Furthermore, we show that there are tractable problems, such as \problem{Small Connected Vertex Cover}, for which one cannot hope to obtain similar results.
\end{abstract}

\section{Introduction}

In many applications we are faced with inputs that are partially uncertain or incomplete. For example, a crucial part of the input, like availability of particular machines or current congestion of some network or road links, may only be available at short notice and may be subject to frequent change. Similarly, we may have to regularly solve instances of some problem that are very similar except for small modifications, e.g., airport gate scheduling when there are only few irregular flights.

A natural approach to this is to come up with solutions that are robust in the sense that they are close to optimal no matter what instantiation the unknown or uncertain part takes. Intuitively it is clear that one cannot hope to always find a solution that is optimal for all instantiations since then the missing parts would always need to be irrelevant. Similarly, if one has to commit to some solution containing uncertain weights/values, then changing these values can in general rule out any good ratio of robustness.

To avoid this issue, in the present work, when given an instance with missing or uncertain information we do not seek to already commit to a solution but to determine how much of the input we can solve or preprocess without knowing the missing or uncertain parts. In particular, this approach permits us to still perform computations once the entire input is known/certain. Thus, there is no general argument that would rule out the possibility of finding optimal solutions since we could always do nothing and just keep the instance as is. 

Our question is rather, assuming that we always want to get an optimal solution, how much of the certain part of the input do we need to keep (including any other derived information that we could choose to compute). Clearly, we should expect that increasing the amount of uncertain data should drive up the amount of information that we need to keep. Conversely, in many settings the instantiations of some $k$ bits (e.g. presence of certain $k$ edges in a graph) ``only'' create $2^k$ different possible instances. Thus, size exponential in the amount of uncertain data is likely to be easy to achieve, e.g., by hardwiring optimal solutions for all instantiations. In contrast, we are interested in spending only polynomial time (too little to precompute an exponential number of solutions) and preprocessing to a size that is polynomial in the amount of uncertain data.

As an easy positive example, consider a road network modeled by a graph $G=(V,E)$ with weights $w\colon E\to\R_{\geq 0}$ capturing the time to travel along the corresponding road. If all weights are given/certain, then we can easily compute a shortest $s$,$t$-path. If, say, weights for edges in some set $F\subseteq E$ are not known (yet) or if they are subject to change (e.g. by road congestion) then we cannot for sure determine a shortest path. Moreover, we cannot in general find a path that will be within any bounded factor of the shortest path: If we have to pick one of two parallel edges, then letting the other one have cost $\epsilon$ and ours have cost $1$ gives ratio $1/\epsilon$ for arbitrary small $\epsilon$. Preprocessing for this setting, however, is straightforward: The final shortest path will consist in some arbitrary way of edges in $F$ and shortest $u$,$v$-subpaths containing no edge of $F$ for $u,v\in\{s,t\}\cup V(F)$. The required $u$,$v$-paths can be precomputed by taking shortest paths in $G-F$. All distance information can be stored in a smaller graph on vertex set $\{s,t\}\cup V(F)$ by letting weight of $\{u,v\}$ be equal to the length of a shortest $u$,$v$-path in $G-F$. The edges of $F$ are then additional parallel edges and the actual shortest $s$,$t$-path can be computed once their weights are known. (One may label edges $\{u,v\}$ by the interior vertices of the shortest $u$,$v$-path to quickly extract a shortest $s$,$t$-path.) Thus, instead of having to find a shortest path in $G=(V,E)$ once all weights are known it suffices to solve the problem on a graph with at most $2+2|F|$ vertices.

\textbf{Our results.}
We study similar preprocessing questions for several fundamental problems, namely \mst, \mwmb, and \mcbm. In the first two problems, the uncertainty lies in the weight of some of the edges of the input graph respectively elements of the ground set of the matroid. The matroid basis problem of course generalizes the \mst question but the latter is probably more accessible and uses essentially the same ideas. For \mcbm we study the setting that in the given bipartite graph $G=(L,R;E)$ there are sets of vertices $L_0\subseteq L$ and $R_0\subseteq R$ some of which will not be available (but we do not know yet). To some extent, the latter problem can also be handled by the result for \mwmb, but the output would not be an instance of bipartite matching (see end of Section~\ref{section:matroidbasis}). For all three problems, we give efficient algorithms that derive an appropriate form of equivalent instance such that an optimal solution can be found using just this instance plus the missing input data. Finally, we show that there are problems for which we cannot find efficient compressions that capture all possible scenarios, even if the running time of the algorithm is allowed to be unbounded; these are \problem{Small Connected Vertex Cover} and some LP-related problems.

\textbf{Related work.}
The effect of uncertainty in instances on optimal solutions has long been considered and there are several approaches to deal with it. An early approach is \emph{stochastic optimization} (SO), starting at least from Dantzig's original paper \cite{Dantzig04}, which assumes that the uncertainty has a probabilistic description. A more recent approach to optimization under uncertainty is \emph{robust optimization} (RO). In contrast to SO, the uncertainty model in RO is not stochastic, but rather deterministic and set-based. More precisely, in RO, we want to find a solution to optimize the value of a \emph{certain} function which subjects to a list of \emph{uncertain} constraints. Each of these constraints may depend on some uncertain parameters whose values are in a given domain which may be infinitely large or continuous. The solution is required to satisfy every constraint for \emph{all} values of its uncertain parameters. We refer interested readers to a survey \cite{BertsimasBC11} for more information about RO. Both of the two approaches (SO and RO) only work with the uncertainty of the \emph{value} of parameters in an instance, but neither with the uncertainty of \emph{appearance} of any factor in an instance (variables, constraints, vertices, edges, etc.), nor with the uncertainty of the objective function. Moreover, the focus is on finding optimal or approximate solutions rather than preprocessing. 

Concerning preprocessing, a related concept is that of \emph{kernelization} from parameterized complexity. In brief, a kernelization for a \emph{decision} problem is an efficient algorithm that compresses each input instance to an instance with smaller size (if possible) and ensures that the answer does not change. Note that the target of this approach is \NP-hard problems and sizes of compressed instances are measured by some problem-specific parameter instead of the input size since one cannot hope to efficiently shrink all inputs of some \NP-hard problem, unless $\P=\NP$. To the best of our knowledge there has not been much research in this area regarding uncertainty or robustness. Two recent results on kernelization nevertheless use intermediate results that are in line with the present work, and that have in part inspired it:

(1)~A nice result of Pilipczuk et al.~\cite{PilipczukPSL14} is the following: Given a plane graph $G$ with outer face $B$, one can efficiently compress the inner part of $G$ to obtain a smaller graph $H$ such that $H$ contains an optimal Steiner tree connecting terminal set $S$ for \emph{every} subset $S\subseteq B$ of the outer face. Note that for any fixed set $S\subseteq B$ this is a polynomial-time problem, but there are of course $2^{|B|}$ many possible sets $S$. Pilipczuk et al.\ use this result to obtain polynomial kernels for several problems on planar graphs, e.g., \problem{Planar Steiner Tree}.

(2)~Kratsch and Wahlstr\"om~\cite{KratschW12} obtained the following result on cut-covering sets: Given a graph and two vertex sets $S$ and $T$, one can efficiently compute a small vertex set $Z$ such that $Z$ contains a minimum vertex $(A,B)$-cut for \emph{every} $A \subseteq S$ and $B \subseteq T$. Again, for each choice of $A$ and $B$ this is polynomial-time solvable, but there is an exponential number of choices. We will make use of this result in Section~\ref{section:matching} and, furthermore, it directly yields another positive example for the \problem{Min Cut} problem with uncertain vertices: Given a graph $G=(V,E)$ with two vertices $s,t \in V$ and $U \subseteq V \setminus \{s,t\}$, we can apply the result for $S = U \cup \{s\}$ and $T = U \cup \{t\}$ to compute a cut-covering set $Z$. Now, for every $U' \subseteq U$, the set $Z$ contains a minimum $U' \cup \{s\}$, $U' \cup \{t\}$ cut $C$. Clearly, $U'\subseteq C$ and thus $C\setminus U'$ must be a minimum $(s,t)$-cut in $G-U'$. This observation together with a technique called \emph{torso operation} (see \cite{MarxOR13}), which will be explained in Section \ref{section:matching}, allows us to compress the certain part of $G$ efficiently. By Menger's theorem, this carries over to the problem of computing the maximum number of vertex-disjoint paths between two vertices in a graph. Thus, one can also obtain a preprocessing for \problem{Max Flow} when all capacities are small, in the sense that the guaranteed size depends polynomially on the maximum capacity. 

Other somewhat related paradigms are \emph{online algorithms} and \emph{dynamic algorithms}. In the former, the input is revealed piece-by-piece and the algorithm needs to commit to decisions without knowing the remaining input; the goal is to optimize the ratio between the online solution and an offline optimum. This is quite different from our setting because it requires to commit to a solution. In the dynamic setting a complete instance is given but modifications to it are given in further rounds; the goal is to adapt quickly to the modifications and to find a solution for the modified instance (faster than computing from scratch). This is closer to our setting, by allowing a different solution in each round, but differs by having a complete input in each round and not necessarily restricting the parts of the input that may change. 

The problem of finding a minimum spanning tree when edge weights are uncertain has also been explored in a different setting \cite{hoffmann_et_al:LIPIcs:2008:1358, ESA:MST}. In this case, all edge weights fall in prespecified intervals and there is a cost for finding out the weight of a specific edge. The goal is to find a cost-efficient query strategy for finding a minimum spanning tree.

\textbf{Organization.}
We will start with preliminaries in Section~\ref{section:prelim} and consider \mst in Section~\ref{section:MST} as a warm-up. We present our algorithms for \mwmb and \mcbm in Sections~\ref{section:matroidbasis} and~\ref{section:matching} respectively. Finally, we present our lower bounds in Section~\ref{section:LB}.

\section{Preliminaries} \label{section:prelim}
\textbf{Graphs.} We mostly follow graph notation as given by Diestel~\cite{Diestel2005graph}. A \emph{walk} in a graph $G$ is a sequence of vertices $(v_0, v_1,\ldots, v_k)$ such that for every $i = 0, \ldots,k-1$ the vertices $v_i$ and $v_{i+1}$ are adjacent; if $G$ is a directed graph then it is required that there is an arc directed from $v_i$ to $v_{i+1}$. A \emph{path} in $G$ is a walk $(v_0, v_1,\ldots, v_k)$ such that $v_i$ and $v_j$ are distinct for every $i \neq j$. In this case, $v_0$ and $v_k$ are called the first vertex and the last vertex of the path respectively; all other vertices are called \emph{internal vertices}. A \emph{$(u,v)$-path} is a path whose first vertex is $u$ and whose last vertex is $v$. If $S$ is a vertex set of a graph $G$, then we denote by $G - S$ the graph obtained from $G$ by removing all vertices in $S$ and their incident edges.

\textbf{Matroids.} \emph{A matroid} is a pair $(E, \I)$, where $E$ is a finite set of elements, called \emph{ground set}, and $\I$ is a family of subsets of $E$ which are called \emph{independent sets} such that:
\begin{inparaenum}[(1)]
 \item $\emptyset \in \I$.
 \item If $A \in \I$, then for every subset $B \subseteq A$ we have $B \in \I$.
 \item If $A$ and $B$ are two independent sets in $\I$ and $|A| > |B|$, then there is an element $e \in A \setminus B$ such that $B \cup \{e\} \in \I$; this is called the \emph{augmentation property}.
\end{inparaenum}
By the augmentation property, all (inclusion-wise) maximal independent sets have the same cardinality; each of them is called \emph{a basis}. The (inclusion-wise) minimal dependent sets are called \emph{circuits}. 
Given a matroid $\M = (E, \I)$ and $F \subseteq E$, we denote by $\M - F$ the matroid obtained from $\M$ by deleting elements in $F$, i.e., the matroid on ground set $E \setminus F$ whose independent sets are the independent sets of $\M$ that are disjoint from $F$. If $F$ is an independent set of $\M$ then we denote by $\M / F$ the matroid obtained from $\M$ by contracting $F$, i.e., the matroid on ground set $E \setminus F$ such that a set $I$ is independent if and only if $I \cup F$ is an independent set of $\M$. A matroid $\M'$ obtained from $\M$ by a sequence of deletion and contraction operations is called a \emph{minor} of $\M$.

A matrix $M$ over a field gives rise to a matroid $\M$ whose ground set is the set of columns of $M$ and a set of columns is an independent set of $\M$ if and only if it is linearly independent as a set of vectors. In this case, we say that $\M$ is \emph{represented by} $M$ or $M$ is a \emph{representation matrix} of $\M$. Note that there may be different representation matrices of the same matroid and there exist matroids which cannot be represented over any field. 

Because there can be as many as $2^{|E|}$ independent sets in a matroid $\M=(E,\I)$, to achieve time polynomial in $|E|$ it is necessary to use a more succinct representation, rather than listing all sets in $\I$ explicitly. Two common ways are representing $\M$ by a matrix (not possible for all matroids) or assuming that an independence oracle for $\I$ is provided:
\begin{description}
 \item[(i)] If our matroid is given by a representation matrix over some field, 
the output of our algorithm should again be a representation matrix. It is known that a representation for any minor of $\M$ can be computed in polynomial time from a representation of $\M$ (cf.~Marx \cite{Marx09}).
 \item[(ii)] If our matroid is given by a ground set and an independence oracle, i.e., a blackbox algorithm which tells us whether an arbitrary subset of the ground set is independent or not, then the time for the oracle is not taken into account. In this case, the output should again be a ground set together with an oracle. Since the oracle is blackbox, the output oracle will be a frontend to the initial oracle and make queries to it.
\end{description}

\textbf{Further notation.} Given two functions $f_1\colon X_1 \to \N$ and $f_2\colon X_2 \to \N$ with $X_1 \cap X_2 = \emptyset$, the \emph{union} of $f_1$ and $f_2$, denoted by $f_1 \cup f_2$, is the function $f \colon X_1 \cup X_2 \to \N$ defined by $f(x) = f_i(x)$ if $x \in X_i$ for $i=1,2$. For convenience, we use $+$ and $-$ instead of $\cup$ and $\setminus$ for singleton sets, e.g., $S + e$ and $S - e$ instead of $S \cup \{e\}$ and $S \setminus \{e\}$. We also abuse notation by using $f(e)$ to represent a function $f$ whose domain is $\{e\}$, e.g., we write $f(e) \cup g$ to clarify that we take the union of two functions $f$ and $g$ where the function $f$ has a singleton domain.

\section{Minimum spanning tree in graphs with some unknown weights}\label{section:MST}

For a connected graph $G=(V,E)$ and a weight function $\omega\colon E\to \N$ one can efficiently compute a spanning tree of minimum total edge weight. If, however, the weight of some edges is not known (yet), then in general we cannot (yet) solve the instance. Nevertheless, we may be able to preprocess the known part of the instance in order to save time later. 

Say we are given a connected graph $G=(V,E\cup F)$ and a weight function $\omega_E\colon E\to \N$. Over different choices of weights $\omega_F\colon F\to\N$ for edges in $F$ there may be an exponential number of different minimum weight spanning trees. We will show how to efficiently generate a new instance on which we may solve the problem for any instantiation of $\omega_F$, i.e., computing the weight of a minimum spanning tree relative to weights $\omega=\omega_E\cup \omega_F$.

In slight abuse of notation we assume that edges that are originally in $F$ can be identified in the new instance after the operations (edge contractions) performed in our algorithm. That is, given an edge in $F$, we can find the corresponding edge in the new instance even if the endpoints of this edge have changed. More formally this could also be captured by an appropriate bijection from $F$ to a set of edges $F'$ that appear in the new instance.

\begin{theorem} \label{thm:mwst}
There is a polynomial-time algorithm that, given a connected graph $G=(V,E\cup F)$ and a weight function $\omega_E\colon E \to \N$, computes a connected graph $G' = (V', E' \cup F)$ with $|E'| \leq |F|$,
a weight function $\omega_{E'} \colon E' \to \N$, and $k \in \N$, such that, for any $\omega_F\colon F\to\N$, the graph $G$ has minimum spanning tree weight $l$ relative to $\omega_E\cup\omega_F\colon E\cup F\to \N$ if and only if $G'$ has minimum spanning tree weight $l'=l-k$ relative to $\omega_{E'}\cup \omega_{F}\colon E'\cup F\to \N$.
\end{theorem}

Let $\MSF$ denote a minimum weight spanning forest of $G-F$ and let $G_1 = (V, \MSF \cup F)$. The following lemma shows that the weight of a minimum spanning tree in $G_1$ is equal to that of a minimum spanning tree in $G$ for any weight function on $F$.

\begin{lemma} \label{lem:forest}
 For any weight function $\omega_F\colon F\to \N$, there is a minimum spanning tree $\MST_{\omega_F}$ in $(G, \omega_E \cup \omega_F)$ such that $\MST_{\omega_F} \subseteq \MSF \cup F$.
\end{lemma}

\begin{longproof}
 We give a proof by contradiction. Suppose that there is a $\omega_F$ such that there is no minimum spanning tree of $(G, \omega_E \cup \omega_F)$ that uses only edges in $\MSF \cup F$. Let $\MST_{\omega_F}$ denote a minimum spanning tree of $(G, \omega_E \cup \omega_F)$ for which $|\MST_{\omega_F} \setminus (\MSF \cup F)|$ is minimal. By assumption we have that $\MST_{\omega_F} \setminus (\MSF \cup F) \neq \emptyset$; let $e = \{u, v\} \in \MST_{\omega_F} \setminus (\MSF \cup F)$. We have that $e \in \MST_{\omega_F}$, $e \notin \MSF$, and $e \notin F$. Thus, there is a cycle $C$ in $\MSF + e$ that contains $e$. For any $e' \in C$ we have that $\omega(e') \leq \omega(e)$. Otherwise, $\MSF$ is not a minimum weight spanning forest. 
 
 Removing $e$ from $\MST_{\omega_F}$ yields two connected components. There must be another edge $e'$ in $C$ between these components since $C$ is a cycle. If $\MST_{\omega_F}$ already contains $e'$, then it is not a spanning tree. Therefore, we can obtain another spanning tree $\MST_{\omega_F}-e+ e'$. Since $\omega(e') \leq \omega(e)$ we have that this spanning tree does not have higher weight than $\MST_{\omega_F}$. Furthermore, it uses fewer edges of $\MST_{\omega_F} \setminus (\MSF \cup F)$. This contradicts that $\MST_{\omega_F}$ is a minimum spanning tree for which $|\MST_{\omega_F} \setminus (\MSF \cup F)|$ is minimal.
\end{longproof}

Accordingly, the first part of the simplification of $(G,\omega_E)$ consists of replacing $G$ by $G_1$ and restricting $\omega_E$ to the edges of $G_1$, i.e., to the edges of $\MSF\cup F$.

Let $\omega_0\colon F \to \N\colon f\mapsto 0$. If an edge $e \in E$ is used by a minimum spanning tree for $(G, \omega_E \cup \omega_0)$ in which all edges in $F$ have zero weight, then, intuitively, using $e$ is also a good choice for spanning trees with other weight functions $\omega_F\colon F\to \N$.

\begin{lemma} \label{lem:zeroweight}
 Let $\MST_{\omega_0}$ a minimum spanning tree for $(G, \omega_E \cup \omega_0)$. For every $\omega_F\colon F\to\N$, there is a minimum spanning tree $\MST_{\omega_F}$ for $(G, \omega_E \cup \omega_F)$ that uses all edges in $\MST_{\omega_0} \setminus F$.
\end{lemma}

\begin{longproof}
 We give a proof by contradiction. Suppose that there is a $\omega_F$ such that there is no minimum spanning tree of $(G, \omega_E \cup \omega_F)$ that uses all edges in $\MST_{\omega_0} \setminus F$. Let $\MST_{\omega_F}$ a minimum spanning tree for which $|(\MST_{\omega_0} \setminus F) \setminus \MST_{\omega_F}|$ is minimal. By assumption we have $(\MST_{\omega_0}\setminus F) \setminus \MST_{\omega_F} \neq \emptyset$; let $e \in (\MST_{\omega_0} \setminus F) \setminus \MST_{\omega_F}$. We have $e \in \MST_{\omega_0} \setminus F$ and $e \notin \MST_{\omega_F}$. Thus, there is a cycle $C$ in $\MST_{\omega_F}+e$ that contains $e$.
 
 Removing $e$ from $\MST_{\omega_0}$ yields two connected components. There must be another edge $e'$ in $C$ between these components since $C$ is a cycle. If $\MST_{\omega_0}$ already contains $e'$, then it is not a spanning tree. Therefore, we can obtain another spanning tree $\MST_{\omega_0}-e+e'$. It follows that $\omega_E(e') \geq \omega_E(e)$ if $e' \in E$, and $\omega_F(e') \geq \omega_0(e') \geq \omega_E(e)$ if $e' \in F$, since otherwise $\MST_{\omega_0}$ is not a minimum spanning tree. But then we can obtain a spanning tree in $(G, \omega_E \cup \omega_F)$ from $\MST_{\omega_F}$ by replacing $e'$ with $e$, that has at most the same weight but uses more edges in $\MST_{\omega_0} \setminus F$. This contradicts that $\MST_{\omega_F}$ is a minimal weight spanning tree for which $|(\MST_{\omega_0} \setminus F) \setminus \MST_{\omega_F}|$ is minimal. 
\end{longproof}

It follows that we can further simplify $G_1$ by contracting edges of $\MST_0 \setminus F$, since for any weight function $\omega_F\colon F\to \N$ there is a minimum spanning tree that uses these edges. 

\begin{lemma} \label{lem:contract}
 Let $\MST_{\omega_0}$ be a minimum spanning tree in $(G, \omega_E \cup \omega_0)$ and let $e \in \MST_{\omega_0} \setminus F$. Let the graph $G_2 = (V', E')$  be obtained by contracting $e$,  where $E'=E\setminus-e$, and let $\omega_{E'}$ be $\omega_E$ restricted to $E'$. For any weight function $\omega_F\colon F\to\N$, $(G, \omega_E \cup \omega_F)$ has a spanning tree with weight $l$ if and only if $(G_2, \omega_{E'} \cup \omega_F)$ has a spanning tree $\MST'_{\omega_F}$ of weight $l-\omega_E(e)$.
\end{lemma}

\begin{longproof}
 By Lemma~\ref{lem:zeroweight}, $(G, \omega_E \cup \omega_F)$ has a spanning tree $\MST_{\omega_F}$ with weight $l$ that uses $e$. Contracting $e$ gives a spanning tree $\MST'_{\omega_F}$ for $(G_2, \omega_{E'} \cup \omega_F)$ with weight $l - \omega_E(e)$. Conversely, if $(G_2, \omega_{E'} \cup \omega_F)$ has a spanning tree $\MST'_{\omega_F}$ with weight $l-\omega_E(e)$, then we obtain a spanning tree $\MST_{\omega_F}$ for $(G, \omega_E \cup \omega_F)$ of weight $l$ by undoing the contraction of $e$ and adding $e$ to the spanning tree.
\end{longproof}

We compute a minimum spanning tree $\MST_{\omega_0}$ for $(G_1, \omega_E \cup \omega_0)$ and create a graph $G' = (V', E' \cup F')$ by contracting every edge in $\MST_{\omega_0} \setminus F$. Let $k$ be the combined weight of the contracted edges. All edges in $E'$ correspond to edges $E$ after the contractions. We define $\omega'_E$ accordingly. 
Note that $|E'| \leq |F|$ since at most $|\MSF| \leq n-1$ edges of $E$ remain in $G_1$, of which we contract at least $|\MST_{\omega_0} \setminus F| \geq|\MST_{\omega_0}| - |F| = n-1-|F|$ edges in order to obtain $G'$.
As a result of Lemmas~\ref{lem:forest} through \ref{lem:contract} we have that $G'$, $\omega'_E$, and $k$ satisfy the required properties in Theorem~\ref{thm:mwst} and the result follows.

\section{Minimum weight basis in matroids with some unknown weights}\label{section:matroidbasis}

Given a matroid $\M=(E,\I)$, we know that for each fixed weight function $w\colon E \to \N$ we can find a minimum weight basis of $\M$ in polynomial time by the greedy algorithm. Now suppose that there is a subset $F \subseteq E$ of elements with unknown weights, i.e., we are only given a partial weight function $w\colon E \setminus F \to \N$. We want to reduce the known part of the input such that when given any weights for elements in $F$ we can compute a minimum weight basis.

In the rest of this section, we prove the following result:

\begin{theorem}\label{thm:matroid}
 There is a polynomial-time algorithm that given a matroid $\M=(E,\I)$, by matrix representation or independence oracle, together with a set $F \subseteq E$ and a partial weight function $w\colon E \setminus F \to \N$, outputs a matroid $\M'= (E', \I')$, a partial weight function $w'\colon E' \setminus F \to \N$, and a number $k \in \N$ such that:
 \begin{inparaenum}[(1)]
  \item $E'$ contains $F$ and has at most $2|F|$ elements.
  \item For every weight function on $F$, the matroid $\M$ has a minimum weight basis of weight $l$ if and only if $\M'$ has a minimum weight basis of weight $l - k$.
  \end{inparaenum} 
\end{theorem}

We start by recalling some basics about the interplay of circuits and bases in a matroid.

\begin{lemma}[cf.\ Oxley {\cite[Corollary 1.2.6]{Oxley-matroidtheory}}]\label{lem:matroid_basis}
 Let $\M = (E, \I)$ a matroid and $B \in \I$ a basis of $\M$. For every $e \in E\setminus B$ there is a unique circuit $C$ in $B + e$, and this circuit contains $e$. Moreover, for every $e' \in C - e$, the set $B + e - e'$ is also a basis of $\M$.
\end{lemma}

\begin{longproof}
 Because $B$ is a basis, $B + e$ must be dependent and therefore there must be a circuit $C$ in $B + e$. The circuit $C$ must contain $e$ because if $e \notin C$, then $C \subseteq B$ and $B$ cannot be an independent set. Hence we know that there must be a circuit $C$ in $B + e$ that contains $e$. Now assume that we have two different such circuits $C_1$, $C_2$. Thus, by basic properties of circuits in matroids, there is a circuit in $(C_1 \cup C_2) - e$. This, however, is impossible since $(C_1 \cup C_2) - e \subseteq B$ is independent and cannot contain any circuit.
 
 Now let $e' \in C - e$. By definition, $C - e'$ is an independent set of size $|C - e'| = |C - e| \le |B|$ (because $C - e \subseteq B$). By the augmentation property of matroids, we can add elements of $B$ to $C-e'$ to obtain an independent set of size $|B|$. This process must end with $B + e - e'$ because we cannot add $e'$ to $C - e'$ to form an independent set. The set $B + e - e'$ is an independent set with the same size as the basis $B$, hence it must also be a basis.
\end{longproof}

The next lemma is about the relation between minimum weight bases of a given matroid $\M$ and the ones in a sub-matroid $\M'$ of $\M$.

\begin{lemma}\label{lem:matroid_delete}
 Let $\M= (E, \I)$ a matroid and let $F \subseteq E$. For every weight function, if $I$ is a minimum weight basis of $\M - F$, then there is a minimum weight basis of $\M$ that is contained in $I \cup F$.
\end{lemma}

\begin{longproof}
 We first prove the lemma for the case $|F| = 1$ and then use induction. Concretely, the base case establishes the lemma for every matroid $M=(E,\I)$ and every $F\subseteq E$ with $|F|=1$. Assume that $F =  \{e\}$ and $I'$ is a minimum weight basis of $\M$ such that $|I' \setminus (I \cup F)|$ is minimum. There are two cases:
 
\textbf{Case 1: $|I'| = |I| + 1$.} By the augmentation axiom, there is an element $e' \in I' \setminus I$ such that $I + e'$ is an independent set in $\M$. Because $I$ is a basis of $\M - F$, $e'$ cannot belong to $\M - F$ (otherwise $I + e'$ is an independent set in $\M - F$ that is larger than $I$). Hence, $e'$ must be $e$ and $I + e$ is a basis in $\M$. On the other hand, because $e = e' \in I'$, $I' - e$ is an independent set in $\M - F$ with the same size as $I$, hence it is also a basis in $\M - F$ and therefore $w(I' - e) \ge w(I)$. We have $w(I + e) = w(I) + w(e) \le w(I' - e) + w(e) = w(I')$. Thus $I + e$ is a minimum weight basis in $\M$ and it is contained in $I\cup F$, as claimed.

\textbf{Case 2: $|I'| = |I|$.} If $e \notin I'$, then $I'$ is also a basis in $\M - F$ and therefore $w(I') \ge w(I)$ which implies that $I$ is also a minimum weight basis in $\M$; in this case we are done since $I\subseteq I\cup F$. Now assume that $e \in I'$. If $I'\subseteq I\cup F$ then we are done, so assume $I' \setminus (I \cup F) \neq \emptyset$ and pick $e' \in I' \setminus (I \cup F)$ with minimum weight; note that $e'\neq e$ because $e'\notin F$. Because $I$ is an independent set with the same size as the basis $I'$, it must be also a basis in $\M$. By Lemma \ref{lem:matroid_basis} there is a unique circuit $C$ in $I + e'$ which contains $e'$ and for every $u\in C-e'$ the set $I+e'-u$ is a basis of $\M$; it is also a basis of $\M-F$ because $I+e'-u\subseteq E\setminus F$. This directly implies that $w(I+e'-u)\geq w(I)$ since the latter is a minimum weight basis of $\M-F$, which entails $w(e')\geq w(u)$ for each $u\in C-e'$. Because $I'$ is an independent set, $C$ cannot be a subset of $I'$ and $C \setminus I'\neq \emptyset$. Pick any $u\in C\setminus I'$ and note that $u\notin\{e,e'\}$ because $e,e'\in I'$. This also means that $u\in C-e'\subseteq I$ and $w(u)\leq w(e')$. Now, by Lemma \ref{lem:matroid_basis} for $I'$ and $u$, there is a unique circuit $C'$ in $I' + u$ and for any $x\in C'-u$ we have that $I'+u-x$ is a basis; clearly $C'-u\subseteq I'$. There are two sub-cases:
\begin{description}
 \item[(i)] We have $C'\nsubseteq I\cup F$; let $e''$ any element in $C'\setminus (I\cup F)$. We get $e''\neq u$ because $u\in I$ and hence $e''\in C'-u\subseteq I'$, implying $e''\in I'\setminus (I\cup F)$. By choice of $e'$ as minimum weight element in $I'\setminus (I\cup F)$ we get $w(e'')\geq w(e')$ and thus $w(e'')\geq w(e')\geq w(u)$. We know that $I'+u-e''$ is a basis, using the above statement for $e''=x\in C'-u$, and it has weight $w(I'+u-e'')=w(I')+w(u)-w(e'')\leq w(I')$. Thus $I'+u-e''$ is also a minimum weight basis but with one less element from $I'\setminus (I\cup F)$, contradicting the choice of $I'$.

 \item[(ii)] We have $C'\subseteq I\cup F= I\cup\{e\}$. It follows that for $C'$ to be a circuit it must contain $e$. Recall that $u\neq e$ so, by Lemma \ref{lem:matroid_basis}, we get that $I'+u-e$ is also a basis of $\M$. It is also a basis of $\M-F$ because it is a subset of $E\setminus F$. Thus, $w(I'+u-e)\geq w(I)$ since the latter is a minimum weight basis for $\M-F$. This directly implies $w(I+e-u)\leq w(I')$. Thus, $I+e-u$ is also a minimum weight basis for $\M$, and it is contained in $I\cup F$, as claimed.
\end{description}
This completes the proof of the lemma for all $\M=(E,\I)$ and $F\subseteq E$ with $|F|=1$.

Now assume that the lemma is true for every matroid $\M'=(E',\I')$ and $F'\subseteq E'$ with $|F'|<k$; we show that it is also true for $\M=(E,\I)$ and $F\subseteq E$ with $|F|=k$. Let $e \in F$, and let $F' = F - e$ and $\M' = \M - F'$. By using the lemma for $\M'$ and $\{e\}$ we get that if $I$ is a minimum weight basis of $\M - F = \M' - e$ then there is a minimum weight basis $I'$ of $\M'$ that is contained in $I \cup \{e\}$. Now we apply the lemma for $\M$ and $F'$ and get that there is a minimum weight basis $I''$ of $\M$ such that $I'' \subseteq I' \cup F' \subseteq I \cup \{ e \} \cup F' = I \cup F$. 
\end{longproof}

Given a matroid with a non-empty subset $F$ of elements, there are infinitely many weight functions on $F$. However, since we are considering a minimization problem, there is a special one: the weight function which assigns value zero to every element in $F$. Intuitively, because elements in $F$ have ''cheapest cost`` in this case, if an element not in $F$ appears in a minimum weight basis with respect to this weight function then it should also appear in some minimum weight basis with respect to other weight functions. The next lemma verifies this intuition.

\begin{lemma}\label{lem:matroid_contract}
Let $\M=(E,\I)$ a matroid, let $F\subseteq E$, and let $\hat{w}\colon E\setminus F\to\N$. If $I_0$ is a minimum weight basis of $\M$ subject to $w_0=\hat{w}\cup \hat{w}_0$ where $\hat{w}_0\colon F\to\N\colon f\mapsto 0$, then for every weight function $w_F\colon F\to\N$ there is a minimum weight basis of $\M$ subject to $w=\hat{w}\cup w_F$ that contains $I_0 \setminus F$.
\end{lemma}

\begin{longproof}
We will prove the lemma for the case that $|F|=1$ and then use induction. Let $F=\{f\}$ and let $w_F\colon F\to\N$. Let $I$ a minimum weight basis of $\M$ corresponding to $w=\hat{w}\cup w_F$ such that $|I \cap I_0 |$ is as large as possible. There are three cases.

If $f \notin I_0$ then we have $w(I_0) = w_0(I_0) \le w_0(I) \le w(I)$, which implies that $I_0$ is also a minimum basis corresponding to $w$; in particular, it contains $I_0\setminus F$, as claimed.

If $f \in I \cap I_0$ then, using that $w_0(f) = 0$ and $w(e) = w_0(e)$ for every $e \neq f$, we have:
\[w(I_0) = w_0(I_0) + w(f) \le w_0(I) + w(f) = w(I),\]
which also implies that $I_0\supseteq I_0\setminus F$ is a minimum basis corresponding to $w$.

The final case is that $f \in I_0 \setminus I$; we show that $(I_0 - f) \subseteq I$ by contradiction. Assume that $(I_0 -f) \setminus I \neq \emptyset$ and pick $e_0$ in $(I_0 -f) \setminus I$ with minimum weight. By applying Lemma \ref{lem:matroid_basis} for $I$ and $e_0$ we get that there is a circuit $C$ in $I + e_0$ which contains $e_0$. Because $C$ cannot be a subset of $I_0$, the set $C \setminus I_0$ is non-empty. We have $w(e) < w(e_0)$ for every element $e \in C \setminus I_0$: Assume that there is an element $e \in C \setminus I_0\subseteq C-e_0$ such that $w(e) \ge  w(e_0)$. Thus, by Lemma \ref{lem:matroid_basis},the set $I' = I + e_0 - e$ is also a basis in $\M$ and it has weight $w(I') = w(I) + w(e_0) - w(e) \le w(I)$. Since also $|I' \cap I_0| > |I \cap I_0|$, this contradicts the way we choose $I$. Thus, $w(e) < w(e_0)$ for every element $e \in C \setminus I_0$.

We first show that $|C \setminus I_0| = 1$. Suppose that there are two different elements $e_1, e_2 \in C \setminus I_0$. For each $i=1,2$, by Lemma \ref{lem:matroid_basis} for $I_0$ and $e_i$, there must be a circuit $C_i$ in $I_0 + e_i$ which contains $e_i$. We know that $C_i$ must contain at least one element not in $I$. If $C_i$ contains an element $u \in (I_0 - f) \setminus I$ then by choice of $e_0$ we have $w(u) \ge w(e_0) > w(e_i)$. On the other hand,  by Lemma \ref{lem:matroid_basis}, $I'_0 = I_0 + e_i - u$ is also a basis, and because $w_0(u) = w(u)$ (note that $u \notin F$) and $w(e_i) \ge w_0(e_i)$, we have $w_0(I'_0) = w_0 (I_0) + w_0(e_i) - w_0(u) \le w_0(I_0) + w(e_i) - w(u) < w_0(I_0)$, which contradicts to the choice of $I_0$. Thus, $C_i$ contains no element of $(I_0-f)\setminus I$ but since $C_i\subseteq I_0+e_i$ this only leaves $e_i$ or $f$ as possible elements in $C_i\setminus I$. However, $e_i\in I$ because $e_i\in C\setminus I_0\subseteq C-e_0\subseteq I$ and, thus, we must have $C_i\setminus I=\{f\}$. Hence, both $C_1$ and $C_2$ are circuits in $I+f$ but then, by Lemma \ref{lem:matroid_basis}, we get $C_1=C_2$. This, however, implies that $C_1-e_1-f=C_2-e_1-f$, which contains $e_2\notin I_0$ and cannot be a subset of $I_0$, contradicting $C_1-e_1\subseteq I_0$. (Note that $f\neq e_i$ because $e_i\in C\setminus I_0$ and $f\in I_0$.)

Now we have $|C \setminus I_0| = 1$, say $C \setminus I_0 = \{e\}$; note that $e\neq f$ because $f\in I_0$. Thus $C - e \subseteq I_0$. Hence $C$ is a circuit in $I_0 + e$ and therefore, by Lemma \ref{lem:matroid_basis}, we have that $I'_0 = I_0 + e - e_0$ is a basis, and it has weight $w_0(I'_0) = w_0(I_0) + w_0(e) - w_0(e_0) = w_0(I_0) + w(e) - w(e_0) < w_0(I_0)$, using that $e\neq f$ and $e_0\neq f$ because $e_0\in (I_0-f)\setminus I$, which contradicts to the way we choose $I_0$. This contradiction also finishes the proof for the case $|F| = 1$.

Now assume that the lemma is true for every $F' \subseteq E$ with $|F'| < k$ and partial weight function $\hat{w}'\colon E \setminus F' \to \N$; we show that it is also true for $F\subseteq E$ with $|F|=k$ and every partial weight function $\hat{w} \colon E \setminus F \to \N$. Fix a partial weight function $\hat{w}\colon E \setminus F \to \N$ and a weight function $w_F \colon F \to \N$, pick an arbitrary element $e \in F$ and let $F' = F \setminus \{e\}$, we denote by $w_{F'}$ and $\hat{w_0}'$ respectively the restrictions of $w_F$ and $\hat{w_0}$ to $F'$; and $\hat{w}'$ the extension of $\hat{w}$ to $E \setminus F'$ by assigning $\hat{w}'(e) = 0$. Let $I_0$ be a minimum weight basis with respect to $w_0 = \hat{w} \cup \hat{w_0} = \hat{w}' \cup \hat{w_0}'$. Consider $\hat{w}'\colon E \setminus F' \to \N$ as a partial weight function, because the lemma is true for $F'$ and $\hat{w}'$, there is a minimum weight basis $I'_0$ with respect to $\hat{w}' \cup w_{F'} = \hat{w} \cup w_{F'} \cup \hat{w}_0(e)$ which contains $I_0 \setminus F'$. Now consider $\hat{w} \cup w_{F'}\colon E \setminus \{e\} \to \N$ as a partial weight function, because the lemma is true for $\{e\}$ and $\hat{w} \cup w_{F'}$, there is a minimum weight basis $I$ with respect to $\hat{w} \cup w_{F'} \cup w(e) = w$ which contains $I'_0 - e$. Thus, we have $I_0 \setminus F = (I_0 \setminus F')  - e \subseteq I'_0 - e \subseteq I$.
\end{longproof}

Now we describe our algorithm. Given a matroid $\M=(E,\I)$ together with a subset $F \subseteq E$ and a partial weight function $w\colon E \setminus F \to \N$, we compute $\M''$ as follows:
\begin{enumerate}
 \item Compute a minimum weight basis $B$ of $\M - F$.
 \item Compute $\M' = \M[B \cup F]$. If $\M$ is given by a representation matrix $M$, then $\M'$ can be represented by the matrix $M'$ obtained from $M$ by taking only the columns corresponding to the elements in $B \cup F$. If $\M$ is given by an oracle $O$, then an oracle $O'$ for $\M'$ can be obtained easily: Given a set $I$, first check whether $I$ is a subset of $B\cup F$, else return no. Query $O$ for whether $I$ is independent in $\M$ and return the answer.
 \item Compute a minimum weight basis $B_0$ of $\M'$ corresponding to the weight function $w_0\colon E\to\N$ with $w_0(e) = 0$ for all $e \in F$ and $w_0(e)=w(e)$ for $e\in E\setminus F$.
 \item Compute $\M'' = \M' / (B_0 \setminus F)$ and $k = w(B_0 \setminus F)$. If $\M'$ is represented by a matrix $M'$, then a matrix representation for $\M''$ can be derived from $M'$ in polynomial time (cf.~\cite{Marx09}). If $\M'$ is given by an oracle $O'$, then an oracle $O''$ for $\M''$ can be obtained as follows:
 Given a set $I$ first check whether $I$ is a subset of the ground set of $\M''$, else return no. Query $O'$ for whether $I\cup (B_0\setminus F)$ is independent in $\M'$ and return the answer.
\end{enumerate}
It is easy to see that our algorithm runs in polynomial time. Because $B_0$ is a basis in $\M'$ and $B$ is an independent set in $\M'$, we have $B_0 \subseteq B \cup F$ and $|B| \le |B_0|$. The number of elements of $\M'$ not in $F$ is
\[ |B \setminus B_0| \le |(B \cup F) \setminus B_0| = |B \cup F| - |B_0| = |B| + |F| - |B_0| \le |F|.\]

The final lemma, about the correctness of our algorithm, finishes the proof of Theorem~\ref{thm:matroid}.

\begin{lemma} \label{lem:matroid_equal}
For every weight function, the minimum weight of a basis in $\M$ is $l$ if and only if the minimum weight of a basis in $\M''$ is $l - k$.  
\end{lemma}

\begin{longproof}
Fix a weight function $w$ and let  $l$, $l'$, and $l''$ the minimum weight of a basis in $\M$, $\M'$, and $\M''$ respectively. By Lemma~\ref{lem:matroid_delete} for $\M$, we have $l = l'$, and by Lemma~\ref{lem:matroid_contract} for $\M'$, we have $l' = l'' + k$. Hence, we have $l = l'' + k$, as claimed. 
\end{longproof}

Because all bases in a matroid have the same size, the problem of finding a minimum weight basis is equivalent to the problem finding a maximum weight basis: Given a matroid $\M$, for each element $e$ of $\M$, we replace its weight $w(e)$ by $w'(e) = c - w(e)$ where $c$ is a large constant to ensure that all new weights are non-negative. Thus, for every basis $I$ we have $w'(I) = c \cdot |I| - w(I)$ where $c \cdot |I|$ is independent of $I$. Thus, a minimum weight basis with respect to the original weight function must be a maximum weight basis with respect to the new one. However our result cannot be applied freely for the case of maximum weight basis in a matroid with some uncertain weights. In contrast to the minimum weight version where uncertain weights can receive any value but the value must be non-negative, there is no natural upper bound for uncertain weights if we only require them to be non-negative. Hence, a similar result for maximum weight basis with uncertain weights currently requires a fixed upper bound for the value of uncertain weights.

Theorem \ref{thm:matroid} can be applied for several specific matroid classes. For example, Theorem~\ref{thm:mwst} can be obtained by an application to the class of graphic matroids, noting that these are closed under deletion and contraction. We finish this section discussing an application for \emph{transversal matroids}. Given a bipartite graph $G=(L,R;E)$ the transversal matroid $\M=(R,\I)$ has as its independent sets exactly those subsets of $R$ that have a matching into $L$ (equivalently, that are the endpoints of some bipartite matching). If we assign weight $1$ for every element of the matroid then the weight of a maximum weight basis in the matroid is the size of a maximum matching in the original bipartite graph. Uncertain vertices in $R$ can be easily simulated by making their weights be uncertain and using weight $0$ to mean that they are not available. Observe that we have an upper bound for uncertain weights, so by the above arguments, we can apply the result for maximum weight basis to compress our uncertain instance for maximum matching in bipartite graph.

Uncertain vertices in $L$ can be handled by giving each a private neighbor that has uncertain weight: We can set these weights very high (and adjust the target weight of the basis that we are looking for) to enforce that the corresponding uncertain vertex is used to match the private neighbor, thereby preventing a matching with other $R$ vertices. If the vertex is available then set the weight of this private neighbor to 0. 
However, the output would not be an instance of bipartite matching. Unlike graphic matroids, transversal matroids are not closed under contraction (which would give the larger class of gammoids), and thus it seems unlikely that one could directly extract an appropriate graph. We address this by studying the bipartite matching problem with uncertain vertices directly in the following section, using other techniques.

\section{Maximum matching in bipartite graphs with uncertain vertices}\label{section:matching}

Given a bipartite graph $G=(L,R;E)$, a maximum matching of $G$ can be found in polynomial time. Now suppose that there are vertex subsets $L_0\subseteq L$ and $R_0\subseteq R$ and some arbitrary vertex sets $L'\subseteq L_0$ and $R'\subseteq R_0$ may not be available in the final input, i.e., we will be asked for a maximum matching in $G-(L'\cup R')$. Thus, there are $2^{|L_0|+|R_0|}$ possible instances. How much can we simplify and shrink $G$ when knowing only $L_0$ and $R_0$, but not $L'$ and $R'$?
We show that, despite the exponential number of possible final instances, a graph $G'$ with polynomial in $|L_0|+|R_0|$ many vertices is sufficient.

\begin{theorem}\label{thm:matching}
There is a randomized polynomial-time algorithm that, given a bipartite graph $G=(L,R;E)$, $L_0\subseteq L$, and $R_0\subseteq R$, returns a bipartite graph $G'$ with $\Oh((|L_0| + |R_0|)^4)$ vertices and $k\in \N$ such that for any $L' \subseteq L_0$ and $R' \subseteq R_0$, the graph $G-(L' \cup R')$ has maximum matching size $l$ if and only if $G'-(L' \cup R')$ has maximum matching size $l' = l-k$.
\end{theorem}

Let us first recall the concept of \emph{augmenting paths}.

\begin{definition}
 Let $M$ a matching in a $G$. An $M$-augmenting path is a path in $G$ s.t.
 \begin{description}
  \item[(i)] the first and last vertices of the path are not incident to any edge in $M$ and
  \item[(ii)] edges on the path are alternatingly in $M$ and not in $M$.
 \end{description}
\end{definition}

It is well known that if an augmenting path $P$ exists, then we can obtain a matching from $M$ that has one more edge by replacing in $M$ the matched edges on $P$ with the non-matched edges on $P$. This extends in a natural way to packings of vertex-disjoint augmenting paths.

\begin{lemma} \label{lem:disj}
Let $G$ be a graph, let $M$ be any matching in $G$, let $M_0$ be a maximum matching in $G$, and let $r$ denote the maximum number of vertex-disjoint $M$-augmenting paths in $G$. We have that $r = |M_0| - |M|$. 
\end{lemma}

\begin{longproof}
First, assume that there are $r$ vertex-disjoint $M$-augmenting paths $P_1, \ldots, P_r$ in $G$. It follows that we can construct a matching of size $|M| + r$ from $M$ by augmenting along each path $P_i$. This can be done independently since the paths are vertex-disjoint. Since, $M_0$ is a maximum matching we have that $|M_0| \geq |M| + r$ and thus $|M_0| - |M| \geq r$.

Now, let $F = M_0 \Delta M$. Since each vertex is incident to at most one edge of each matching we have that $F$ corresponds to a union of vertex-disjoint paths and cycles in which edges of $M_0$ and $M$ appear alternatingly. On each of the cycles the number of edges of $M_0$ and $M$ are equal; the paths can be partitioned into two types:
\begin{description}
 \item[(i)] Paths with more edges of $M_0$ than edges of $M$.
 \item[(ii)] Paths with at least as many $M$ edges as $M_0$ edges.
\end{description}
Note that paths of type (i) have exactly one more edge of $M_0$ than of $M$ and each such path is an $M$-augmenting path. Because all edges of $M$ and $M_0$ appear in a path or cycle we get that there must be at least $|M_0|-|M|$ paths of type (i) since these are the only structure with (one) more $M_0$ than $M$ edges. All these paths are vertex-disjoint, which implies that $r\geq |M_0|-|M|$.
It follows that $r = |M_0| - |M|$, as claimed.
\end{longproof}

We now fix a maximum matching $M$ in $G- (L_0 \cup R_0)$ and use it in the remainder of the section. We direct the edges of $G$ to obtain a directed bipartite graph $H = (L, R; A)$ as follows: Every edge in $M$ is directed from $R$ to $L$ and every edge not in $M$ is directed from $L$ to $R$. This type of directed graph is part of a folklore approach for finding augmenting paths. Let $F_L$ (resp. $F_R$) denote the vertices of $L \setminus L_0$ (resp. $R \setminus R_0$) which are not covered by $M$, and note that $V(M)$, $L_0$, $R_0$, $F_L$, and $F_R$ are pairwise disjoint. 

\begin{observation} \label{obs:corr}
 For any $L' \subseteq L_0$ and $R' \subseteq R_0$, there is a one-to-one correspondence between directed paths in $H-(L' \cup R')$ from $F_L \cup (L_0 \setminus L')$ to $F_R \cup (R_0 \setminus R')$ and $M$-augmenting paths in $G-(L' \cup R')$. This is because $F_L \cup (L_0 \setminus L')$ and $F_R \cup (R_0 \setminus R')$ are exactly the vertices that are free in the matching, while a directed path must visit matched edges alternatingly, since these are exactly the edges from $R$ to $L$.
\end{observation}

\begin{observation} \label{obs:max}
 For any $L' \subseteq L_0$ and $R' \subseteq R_0$, there is no $M$-augmenting path in $G-(L' \cup R')$ that starts in $F_L$ and ends in $F_R$. This follows from maximality of $M$, since such a path could be used to obtain a bigger matching in $G-(L_0 \cup R_0)$. By Observation~\ref{obs:corr} we have that there is no directed path in $H-(L' \cup R')$ from $F_L$ to $F_R$.
\end{observation}

Given the relation between augmenting paths and directed paths we now consider minimum cuts in the graph $H$. The following theorem of Kratsch and Wahlstr\"om~\cite{KratschW12} provides a small set of vertices that contains minimum cuts for a specified type of requested $A,B$-cuts.

\begin{theorem}[Kratsch and Wahlstr{\"o}m \cite{KratschW12}] \label{thm:cutset} 
 Let $G = (V, E)$ be a directed graph and let $S, T \subseteq V$. Let $r$ denote the size of a minimum $(S,T)$-vertex cut (which may intersect $S$ and $T$). There exists a set $X \subseteq V$ of size $|X| = \Oh(|S| \cdot |T| \cdot r)$ such that for any $A \subseteq S$ and $B \subseteq T$ the set $X$ contains a minimum $(A,B)$-vertex cut. Such a set $X$ can be computed in randomized polynomial time with error probability $\Oh(2^{-n})$.
\end{theorem}

The following lemma adapts Theorem~\ref{thm:cutset} to our application, mainly taking care not to inflate the rank overly much. (We ask for minimum $(F_L \cup (L_0 \setminus L'), F_R \cup (R_0 \setminus R'))$-vertex cuts, but without having rank $\Omega(|F_L|+|F_R|)$.)

\begin{lemma} \label{lem:X}
 There exists a set $X \subseteq L \cup R$ of size $|X| = \Oh((|L_0|+|R_0|)^3)$ such that for~any $L' \subseteq L_0$ and $R' \subseteq R_0$, $X$ contains a minimum $(F_L \cup (L_0 \setminus L'), F_R \cup (R_0 \setminus R'))$-vertex cut~in $H-(L' \cup R')$; it can be found in randomized polynomial time with error probability $\Oh(2^{-n})$.
\end{lemma}

\begin{longproof}
We start by creating an auxiliary directed bipartite graph $H_{aux}$ from $H$ by adding two sets of vertices $\bar{L}$ and $\bar{R}$ of size $|\bar{L}| = |\bar{R}| = |L_0| + |R_0| + 1$. For each vertex $u \in \bar{L}$ and for each vertex $v \in F_L$ we add an arc $(u, v)$. For each vertex $u \in \bar{R}$ and for each vertex in $v \in F_R$ we add an arc $(v, u)$. Let $X$ be obtained by invoking Theorem~\ref{thm:cutset} with $H_{aux}$, $S = \bar{L} \cup L_0 \cup R_0$ and $T = \bar{R} \cup L_0 \cup R_0$. Thus, for any $L' \subseteq L_0$ and $R' \subseteq R_0$, $X$ contains a minimum $(\bar{L} \cup (L_0 \setminus L'), \bar{R} \cup (R_0 \setminus R'))$-vertex cut in $H_{aux}-(L' \cup R')$: To see this, let $A= \bar{L} \cup L_0 \cup R'\subseteq S$ and $B= \bar{R} \cup L' \cup R_0\subseteq T$, and let $C$ a minimum $(A,B)$-vertex cut. Clearly, $A\cap B=L'\cup R'$ must be contained in $C$. Therefore, $C\setminus (L'\cup R')$ must be a $(A\setminus (L'\cup R'), B\setminus (L'\cup R'))$-vertex cut in $H_{aux}-(L'\cup R')$; it must also be minimum or else a smaller cut could be combined with $L'\cup R'$ to get a better $(A,B)$-vertex cut than $C$. Note that $A\setminus (L'\cup R')= \bar{L} \cup (L_0\setminus L')$ and $B\setminus (L'\cup R')=\bar{R} \cup (R_0\setminus R')$.

We show that we may safely remove from $X$ all vertices in $\bar{L} \cup \bar{R}$ without losing any of these cuts: Concretely, we show that each minimum $(\bar{L} \cup (L_0 \setminus L'), \bar{R} \cup (R_0 \setminus R'))$-cut $C$ in $H_{aux}-(L' \cup R')$ must be disjoint from $\bar{L}\cup \bar{R}$. Assume for contradiction that some minimum cut $C$ contains any vertices of $\bar{L}$. It follows that $C$ must contain $\bar{L}$ entirely since these vertices have the same neighborhood. This implies that $|C| \geq |L_0| + |R_0| + 1$. However, by Observation~\ref{obs:max} we have that $H$ does not contain any directed paths from $F_L$ to $F_R$. Thus, by construction of $H_{aux}$ we have that there are no directed paths from $\bar{L}$ to $\bar{R}$. Therefore, we have that $L_0 \cup R_0$ is a $(\bar{L} \cup (L_0 \setminus L'), \bar{R} \cup (R_0 \setminus R'))$-cut of size $|L_0| + |R_0| < |C| \leq |L_0| + |R_0| + 1$, which contradicts the fact that $C$ is a minimum cut.
Note also that this implies $r \leq |L_0| + |R_0|$ when we invoke Theorem~\ref{thm:cutset} and therefore $|X| = \Oh((|L_0| + |R_0|)^3)$.

Hence, we assume that $X \cap (\bar{L} \cup \bar{R}) = \emptyset$. For any fixed $L' \subseteq L_0$ and $R' \subseteq R_0$, if $X$ contains a minimum $(\bar{L} \cup (L_0 \setminus L'), \bar{R} \cup (R_0 \setminus R'))$-cut $C$, then $C$ must be a minimum $(F_L \cup (L_0 \setminus L'), F_R \cup (R_0 \setminus R'))$-cut as well, since $C$ does not contain any vertices from $\bar{L} \cup \bar{R}$ while any $(F_L \cup (L_0 \setminus L'), F_R \cup (R_0 \setminus R'))$-path can be extended to a $(\bar{L} \cup (L_0 \setminus L'), \bar{R} \cup (R_0 \setminus R'))$-path by construction of $H_{aux}$. Thus, if there is a $(F_L \cup (L_0 \setminus L'), F_R \cup (R_0 \setminus R'))$-cut $C'$ that is smaller than $C$ and not contained in $X$, then $C'$ is a smaller $(\bar{L} \cup (L_0 \setminus L'), \bar{R} \cup (R_0 \setminus R'))$-cut as well, which contradicts that $X$ contains a minimum $(\bar{L} \cup (L_0 \setminus L'), \bar{R} \cup (R_0 \setminus R'))$-cut. Therefore, $X$ contains a minimum $(F_L \cup (L_0 \setminus L'), F_R \cup (R_0 \setminus R'))$-cut for any $L'$ and $R'$, which proves the lemma.
\end{longproof}

\begin{definition}
Let $D=(V,A)$ a directed graph and $Z \subseteq V$. By applying to $D$ the \emph{torso operation} on $Z$ we mean to derive a new graph, denoted by $\tor(D,Z)$, by adding to $D[Z]$ an arc $(u,v)$ for every pair $u,v \in Z$ if there is a directed $(u,v)$-path in $D$ with no internal vertices from $Z$. If an arc of $\tor(D,Z)$ is not in $D[Z]$, then we call it a \emph{shortcut arc}.
\end{definition}

We now construct a set $Z$, starting from $X$ as obtained from Lemma \ref{lem:X}: Let $M_X$ be the set of edges in $M$ with at least one endpoint in $X$ and let $X' = X \cup L_0 \cup R_0 \cup V(M_X)$; note that $X'\cap (F_L\cup F_R)=\emptyset$. For each $v \in X' \cap R$ (resp.\ $v \in X' \cap L$), let $F_v \subseteq F_L$ (resp.\ $F_v \subseteq F_R$) denote the set of vertices that can be reached from $v$ (resp.\ can reach $v$) by a directed path in $H$ with no internal vertices from $X'$. If $|F_v| \le |L_0| + |R_0|$, then let $W_v = F_v$; otherwise let $W_v$ be an arbitrary subset of $F_v$ of size $|L_0| + |R_0|$. Finally, let $Z = X' \cup \bigcup_{v \in X'} W_v$.

Let $H' = \tor(H,Z)$. By construction there is no matching edge in $M$ with exactly one vertex in $Z$, which ensures that $H'$ is also a bipartite graph: By construction, a directed path connecting two vertices of the same side must either start or end with an edge in $M$ and therefore that path cannot have only internal vertices in $V \setminus Z$. Thus, the torso operation does not add a shortcut edge between two vertices that are on the same side.

Now let $G'$ be the underlying undirected graph corresponding to $H'$ and let $k = |M[(V \setminus Z]|=|M| - |M_X|$, i.e., the number of edges of $M$ outside of $Z$. 
We show that the maximum matching size of $G-(L'\cup R')$, for $L'\subseteq L_0$ and $R'\subseteq R_0$, can also be computed in $G'$.

\begin{lemma}\label{lem:matching}
 For every $L' \subseteq L_0$ and $R' \subseteq R_0$, the graph $G - (L' \cup R')$ has maximum matching size $l$ if and only if $G' - (L' \cup R')$ has maximum matching size $l - k$.
\end{lemma}

\begin{shortproof}
Let us fix $L' \subseteq L_0$ and $R' \subseteq R_0$ and denote $S = F_L \cup (L_0 \setminus L')$, $T = F_R \cup (R_0 \setminus R')$. 

$(\Rightarrow)$ Let us first assume that $G - (L' \cup R')$ has a maximum matching $M'_0$ of size $l$. By Lemma~\ref{lem:disj} we have that there is a vertex-disjoint packing of $M$-augmenting paths $\Pee$ of size $l-|M|$ which we can use to augment $M$ to a (maximum) matching $M_0$ of size $|M'_0|$. Note that $M_0$ agrees with $M$ except for the augmenting paths in $\Pee$. (The same is not necessarily true for $M'_0$ since there could, e.g., be alternating cycles in $M'_0\Delta M$.) Since every $M$-augmenting path corresponds to a directed path from $S$ to $T$ and $X \subseteq X' \subseteq Z$ contains a minimum $(S, T)$-cut we have that every path in $\Pee$ contains at least one vertex of $Z$. Furthermore $\Pee \leq |L_0| + |R_0$| because every augmenting path must contain at least one vertex of $L_0 \cup R_0$ by $M$ being maximum matching in $G-(L_0\cup R_0)$.

We consider maximal subpaths of $\Pee$ with internal vertices not from $Z$ and having at least one internal vertex. (Note that vertices in $(F_L\cup F_R)\setminus Z\subseteq F_L\cup F_R$ are $M$-unmatched, so paths in $\Pee$ cannot have such vertices as internal vertices: Those in $F_L$ have no incoming edges and those in $F_R$ have no outgoing edges.)
If an internal vertex $v$ on a path in $\Pee$ is in $Z$, then $v \in X' \subseteq Z$ since $Z \setminus X' \subseteq F_L \cup F_R$. If an internal vertex $v$ of a path in $\Pee$ is in $Z$, and thus $v \in X'$, then we have that at least one neighbor of $v$ on the path was also included in $X' \subseteq Z$ since $v$ is incident with $M$. Therefore, these subpaths are vertex-disjoint. Let us consider each such subpath $P = (p, \ldots, q)$. We distinguish three cases, based on whether $p$ and/or $q$ are contained in $Z$, and apply a replacement operation to $M_0$ for each of them. 

If $p, q \in Z$, then we have an edge $\{p, q\}$ in $G'$ because it was either in the original graph, or the corresponding shortcut arc $(p, q)$ was added to $H'$ during the torso operation. Since no edge in $M$ has exactly one vertex in $Z$, we have that $P$ starts and ends with an edge of $M_0$, and there is exactly one less edge of $M$ on $P$ than there are edges of $M_0$ on $P$. We modify $M_0$ by removing all edges of $M_0$ on $P$, adding all edges of $M$ on $P$, and adding $\{p, q\}$. After this modification the size of $M_0$ is the same as before and it is still a matching.

If $p \in Z$ and $q \notin Z$, then we have that $W_p \geq |L_0| + |R_0|$, since otherwise we would have $q \in F_v = W_v \subseteq Z$ since it is reachable from $p$ with no internal vertices in $Z\supseteq X'$. We again have that the first and last edge of $P$ start and end with an edge of $M_0$, since the first edge cannot be an edge of $M$ because it has only one endpoint in $Z$, while the last edge is a final edge of a path in $\Pee$ by maximality of $P$ and $P$ ends in an $M$-unmatched vertex of $F_R$. (It cannot end in $F_L$ because those vertices have no incoming edges.) We modify $M_0$ by removing all edges of $M_0$ on $P$, adding all edges of $M$ on $P$ and adding one edge from $p$ into $W_p$. Because this case can occur at most $|L_0| + |R_0|$ times (at most once for every path in $\Pee$) we have that there is always a free vertex in $W_p$. Again, the size of $M_0$ remains the same.

We handle the case where $p \notin Z$ and $q \in Z$ similarly. Note that $p, q \notin Z$ cannot occur because then $P$ would not a be maximal subpath with internal vertices not from $Z$ since every path in $\Pee$ visits at least one vertex in $Z$. After handling every maximal subpath in this fashion there are no edges of $M_0$ with only one endpoint in $Z$. Furthermore, $M_0$ and $M$ agree on all edges not incident to $Z$ and thus $|M_0[V\setminus Z]| = |M[V \setminus Z]$. Hence, $M_0[Z]$, the restriction of $M_0$ to $Z$, is a matching in $G'-(L'\cup R')$ of size $|M_0[Z]| = |M_0| - |M_0[V \setminus Z]| = l - |M[V \setminus Z]| = l - k$.

$(\Leftarrow)$ Assume that $G' - (L' \cup R')$ has a matching $M_0$ of size $l - k$. Let $M' = M[Z] = M_X$ denote the restriction of $M$ to $Z$. Since there are no edges of $M$ with exactly one vertex in $Z$, the set $M'$ is a matching in $G'-(L' \cup R')$ of size $|M|-k$. By Lemma \ref{lem:disj}, there is a packing $\Pee$ of $r = (l - k) - |M'| = l - |M|$ vertex-disjoint $M'$-augmenting paths in $G' - (L' \cup R')$, which correspond to $r$ vertex-disjoint directed paths from $S \cap Z$ to $T \cap Z$ in $H' - (L' \cup R')$. 

By construction $X$ contains a minimum $(S,T)$-cut $Y$ in $H - (L' \cup R')$; suppose that $|Y| < r=|\Pee|$. There must be a path in $\Pee$ which avoids $Y$. This path corresponds to a directed path $P$ from $S \cap Z$ to $T \cap Z$ in $H' - (L' \cup R')$. Arcs of $P$ are either arcs in $H-(L' \cup R')$ or shortcut arcs which are added by the torso operation. Note that each shortcut arc corresponds to a directed path with no internal vertices from $Z$, which therefore also avoids $Y \subseteq X \subseteq Z$. Hence, if we replace shortcut arcs in $P$ by corresponding paths in $H-(L' \cup R')$, then we obtain a directed walk from $S \cap Z$ to $T \cap Z$ in $H-(L' \cup R')$, which contradicts that $Y$ is a $(S,T)$-cut in $H-(L' \cup R')$. Hence we have $|Y| \ge r$, which implies that there are at least $r$ vertex-disjoint $(S,T)$-paths in $H - (L' \cup R')$. These paths correspond to $M$-augmenting paths in $G - (L' \cup R')$. Thus, $G - (L' \cup R')$ has a matching of size at least $|M| + r = |M| + (l - |M|) = l$.
\end{shortproof}

The size of $X$ is polynomial $\Oh((|L_0| + |R_0|)^3)$. Therefore $|X'| = \Oh((|L_0| + |R_0|)^3)$ since we add at most $|X|$ more vertices that are an endpoint to a matched edge in $M$ incident to $X$. To obtain $Z$, at most $|L_0| + |R_0|$ vertices are added for every vertex in $X'$. Thus, $Z$ is of size $\Oh((|L_0| + |R_0|)^4)$ and therefore $G'$ has at most $\Oh((|L_0| + |R_0|)^4)$ vertices. All operations required to obtain $G'$ can be performed in polynomial time by using appropriate flow calculations. Correctness follows from Lemma~\ref{lem:matching}, completing Theorem~\ref{thm:matching}.

\subparagraph{Uncertain edges.}
Our result can also be extended to the case when there are some \emph{edges} of the input graph that may not be available; we call these edges \emph{uncertain edges}. We proceed as follows: For each uncertain edge $uv$, we subdivide it into three edges $uu', u'v'$ and $v'v$ (note that this does not change the bipartiteness of the graph); then instead of considering $uv$ as an uncertain edge, we consider $u'$ and $v'$ as uncertain vertices. The auxiliary vertices $u'$ and $v'$ can be used to control the availability of $uv$ in the graph. We remove $u'$ and $v'$ to make $uv$ unavailable and include both $u$ and $v$ in the final input to make $uv$ available. Now we can repeat the algorithm in Theorem \ref{thm:matching} with a small modification: Before applying the torso operation, we add $u, u', v'$ and $v$ to the cut-covering set. This ensures that the new edges and vertices that correspond to uncertain edges are not affected by the torso operation. After the torso operation has been applied, we obtain a new compressed graph with some uncertain vertices, and moreover all endpoints of uncertain edges and their subdivisions in the input graph are preserved. Finally, we may contract subdivided edges with auxiliary (uncertain) vertices to get the original uncertain edges. 

The final thing we need to be careful about is how the size of maximum matching may change under subdivision (and contraction). Given a graph $G$, if we subdivide an edge $uv$ of $G$ into three edges $uu', u'v'$ and $v'v$ to obtain a graph $G'$, then we increase the size of maximum matching of $G$ by one. Indeed, we just need to see how to construct a matching in the new graph from a maximum matching $M$ of $G$. If $M$ contains $uv$ then we just need to replace $uv$ by $uu'$ and $vv'$. Otherwise, we add $u'v'$ to $M$; in both cases the size increases by one, as claimed.
Thus, we can conclude our result carries over to the case where for a subset of edges and vertices it is unknown if they appear in the final input.

\section{Lower Bounds} \label{section:LB}

In this section we present some unconditional lower bounds for preprocessing instances of some tractable problems in which part of the input is uncertain. We derive these lower bounds from the {\sc Membership} communication game which is defined as follows. We have two players, Alice and Bob. Alice has a subset $S \subseteq [n]$ and Bob has an integer $l$. The objective of the game is for Bob to find out if $l \in S$. It is well known that a one-way communication protocol from Alice to Bob has a cost of at least $n$ bits of information. We show that for certain problems the existence of algorithms that are similar to those presented in this paper would give a more efficient protocol. We start with {\sc Connected Vertex Cover} which is solvable in time $2^b |V|^{\Oh(1)}$ if we are looking for a solution of size at most $b$ \cite{Cygan12}. Therefore, it is solvable in polynomial time for $b \leq \log |V|$ in which case we refer to the problem as {\sc Small Connected Vertex Cover}. The following theorem shows that we cannot find a succinct equivalent instance if for a set $W$ of vertices it is uncertain if they appear in the final input.

\begin{theorem} \label{thm:scvc}
There is no algorithm that, given an instance $G = (V\cup W, E)$ of {\sc Small Connected Vertex Cover}, outputs a graph $G'$ of size $|W|^{\Oh(1)}$ and $k \in \N$ such that, for any $W' \subseteq W$, the graph $G-W'$ has a connected vertex cover of size at most $b \leq \log |V|$ if and only if $G'-W'$ has a connected vertex cover of size at most $b- k$.
\end{theorem}

\begin{longproof}
For contradiction, let us assume that such an algorithm exists. We show that it could be used to obtain a better protocol for {\sc Membership}. Suppose that Alice has a set $S\subseteq [n]$ and Bob has an integer $l$. For convenience assume that $n=2^p$. 

Alice constructs a graph $G = (V \cup W, E)$ as follows. For every $j \in [n]$ assign a unique identifier of $p = \log n$ bits, denoted by $s(j)$ . Let $V = S' \cup U$, where $S'$ has a vertex $v_j$ for each $j \in S$, and $U$ is of size $2p$. Let $W = W_0 \cup W_1$, where $|W_0| = |W_1| = p$. Let $E$ initially consist of a matching of edges between $U$ and $W_0 \cup W_1$, i.e., each vertex in $W_0 \cup W_1$ has a private neighbor in $U$. For each vertex $v_j \in S'$ corresponding to an element $j \in S$ and $1 \leq i \leq p$, add an edge $\{v_j, w_i\}$, where $w_i$ is the $i$-th vertex of $W_0$ (resp. $W_1$) if the $i$-th bit of $s(j)$ is set to 0 (resp. 1). 

For each element $j \in [n]$ consider the following. Pick a set of vertices $\hat{W}$ of size $p$ such that $w_i$, the $i$-th vertex from $\hat{W}$ is the $i$-th vertex of $W_0$ (resp. $W_1$) if the $i$-th bit of $s(j)$ is set to 0 (resp. 1). Let $W' = W \setminus \hat{W}$. The graph $G-W'$ has a connected vertex cover of size $p + 1$ if and only if $j \in S$: In order to cover the edges between $U$ and $\hat{W}$, we need at least $p$ vertices. Taking any vertex of $U$ forces us to also pick its neighbor in $\hat{W}$ for connectivity since vertices in $U$ have degree~$1$. Thus, any optimal solution takes all vertices in $\hat{W}$. In order to connect these vertices we have a remaining budget of 1. Thus we need a common neighbor of $\hat{W}$. This neighbor exists if and only if $v_j \in S'$ (and  $j \in S$).

Now let us suppose that Alice uses the algorithm to compute $G'$ of size $|W|^{\Oh(1)}$, and $k$. She then sends this to Bob, who can in turn plug in $W'$ corresponding to $s(l)$ and find out if $l \in S$ by testing whether $G'-W'$ has a connected vertex cover of size $p+1-k$. However, the size of the encoding of $G'$ is poly-logarithmic in $n$ since we have $|W| = 2p = 2 \log n$ (and so is the size of the encoding of $k$ since we can assume that $k \leq \log n + 1$). This gives us an $o(n)$ one-way communication protocol for {\sc Membership} which for sufficiently large enough $n$ contradicts the lower bound of $n$ bits.
\end{longproof}

Note that this lower bound holds even if the time to compute $G'$ and $k$ is unbounded. Furthermore, we remark that by a similar information theoretic argument, any encoding from which we can extract the answer to {\sc Small Connected Vertex Cover} for any $W' \subseteq W$ requires at least $2^{|W|/2} = n$ bits: Otherwise, since there are $2^n$ possible subsets of $[n]$, by the pigeonhole principle (we have fewer than $2^n$ distinct bit-strings of length smaller than $n$) there must be at least 2 subsets $S \neq S'$ of $[n]$ for which the encoding of the instance produced in the proof of Theorem~\ref{thm:scvc} is the same. Now there must be an element $i$ that is, w.l.o.g., in $S$ but not in $S'$ for which plugging in the corresponding $W'$ produces the same answer, which is clearly wrong. We show a similar lower bound for {\sc Linear Programming} where for part of the constraints it is uncertain if they appear in the final input.

\begin{theorem} \label{thm:lp}
 There is no algorithm that, given an LP with objective function $\bm{c}^T\bm{x}$ and constraints $A\bm{x} \leq \bm{b}, B\bm{x} \leq \bm{d}, \bm{x} \geq 0$, outputs an encoding of size smaller than $2^{c/2}$ from which we can correctly determine feasibility of the LP for any subset of constraints in $B\bm{x} \leq \bm{d}$, where $c$ is the number of constraints in $B\bm{x} \leq \bm{d}$.
\end{theorem}

\begin{longproof}
 For contradiction, assume that such an algorithm exists. We show how this could be used to violate the lower bound for {\sc Membership}. Suppose that Alice has a set $S\subseteq [n]$ and Bob has an integer $l$. For convenience assume that $n = 2^p$.
 
 Alice creates an LP as follows: Let the objective function be $\bm{0}^T \bm{x}$. Assign for every element $j \in [n]$ a unique identifier of $p = \log n$ bits, denoted by $s(j)$. Let $\bm{x}$ consist of $p$ variables constrained to $0 \leq x_i \leq 1$. Let $A\bm{x} \leq \bm{b}$ consist of a constraint $y_1 + \cdots + y_p \leq p - 0.5$ for each $j \in S$, where $y_i = x_i$ (resp. $y_i = (1-x_i)$) if the $i$-th bit of $s(j)$ is set to 1 (resp. 0). Let $B\bm{x} \leq \bm{d}$ consist of constraints $x_i \leq 0$ and $x_i \geq 1$ for each variable $x_i$ in $\bm{x}$. 
 
 For each element $j \in [n]$ we have the following. Pick a subset of $(B\bm{x} \leq \bm{d})$ by taking for each $x_j$ in $\bm{x}$ the constraint $x_j \geq 1$ (resp $x_i \leq 0$) if the $i$-th bit of $s(j)$ is set to 1 (resp. 0). Together with $0 \leq x_i \leq 1$, for every variable in $\bm{x}$ the only feasible assignment is exactly 0 or 1 according to the corresponding bit in $s(j)$. The LP has a feasible solution if and only if $j \notin S$: For each $j' \in S$, the value of the corresponding $y_1 + \cdots + y_p$ is smaller than $p - 0.5$ if and only if $s(j')$ does not match $s(j)$ in at least one index, allowing the corresponding $y$-term to contribute 0 instead of 1.
 
 Now Alice uses the algorithm and outputs an encoding of size smaller than $2^{c/2}$ and sends this to Bob, who can find out if $l \in S$ by plugging in a subset of constraints $(B\bm{x} \leq \bm{d})$ corresponding to $s(l)$ and finding out, via the encoding, whether the LP constructed by Alice is feasible. However, by the pigeonhole principle (in the constructed instance, $c = 2 \log n$), there must be subsets $S \neq S'$ of $[n]$ for which the output of the algorithm is the same. Thus, the answer to feasibility of the LP that Bob obtains from the outputs for $S$ and $S'$ is the same for any element $l$, even for, w.l.o.g., $l$ in $S$ but not in $S'$. Therefore, the algorithm must be incorrect.
\end{longproof}

Let us remark that a similar lower bound can be obtained for linear programming instances if for part of the constraints the righthand side is unknown, and we would like to obtain an instance that is equivalent with respect to feasibility for any assignment of the righthand side in these constraints. This follows from the proof of Theorem~\ref{thm:lp} when according to $s(i)$, we set $x_i \geq 1$ and $x_i \leq 1$ (resp. $x_i \geq 0$ and $x_i \leq 0$) if the $i$-th bit of $s(i)$ is set to 1 (resp. 0).

Finally, in a setting where the objective function of the LP is unknown we have a lower bound of $2^c$ where $c$ is the number of variables: Alice uses the same constraints $Ax\leq b$ as in Theorem~\ref{thm:lp}. Now, target functions $y_1+\ldots+y_p$ can take value $p$ if and only if there is no constraint $y_1+\ldots+y_p\leq p-0.5$, i.e., if and only if the corresponding element $j$ is not in $S$.

\section{Conclusion}

We have initiated a programmatic study of preprocessing for efficiently solvable problems for the setting that not the entire input is known or that parts of the input are not certain, inspired by recent results \cite{PilipczukPSL14,KratschW12} obtained in the context of kernelization. Intuitively, incomplete or partially uncertain instances correspond to a possibly exponentially large family of similar instances. Thus, it is not clear (and as shown it does not hold in general) that one can efficiently generate an instance of size polynomial in the amount of uncertain data that allows to compute optimal solutions for the initial input for each instantiation (and without access to the initial input of course).
This direction of research seems to apply for a variety of polynomial-time solvable problems (and of course also for \NP-hard ones). In particular, different notions of missing or uncertain data may be suitable for the same problem. We have considered weight functions whose values are unknown for certain items as well as graphs in which some vertices or edges may or may not appear in the instantiation.

Natural problems for future research are for example matchings in general graphs (with uncertain vertices, edges, and or weights) and \problem{Max Flow} with arbitrary and uncertain capacities. It would also be interesting whether there are particular lower bound techniques for this form of preprocessing, e.g., to prove that a certain amount of information is optimal.

\newpage
\bibliography{preprocessing}

\end{document}